\documentclass[aps,prb,twocolumn,showpacs,groupedaddress]{revtex4}
\usepackage{graphicx}
\usepackage{amssymb}
\usepackage{amsmath}
\usepackage[svgnames]{xcolor}
\usepackage{hyperref}
\usepackage{epsf}
\usepackage{psfrag}

\def \be{\begin{equation}}
\def \ee{\end{equation}} 

\begin{document}

\title{Gate-modulated thermopower in disordered nanowires: \\ I. Low temperature coherent regime}
\author{Riccardo Bosisio}
\author{Genevi\`eve Fleury}
\author{Jean-Louis Pichard}
\affiliation{Service de Physique de l'\'Etat Condens\'e (CNRS URA 2464), 
IRAMIS/SPEC, CEA Saclay, 91191 Gif-sur-Yvette, France}

\begin{abstract}
Using a one-dimensional tight-binding Anderson model, we study a disordered nanowire in the presence of an external gate  
which can be used for depleting its carrier density (field effect transistor device configuration). In this first 
paper, we consider the low temperature coherent regime where the electron transmission through the nanowire remains elastic. 
In the limit where the nanowire length exceeds the electron localization length, we derive three analytical expressions for 
the typical value of the thermopower as a function of the gate potential, in the cases where the electron transport takes 
place (i) inside the impurity band of the nanowire, (ii) around its band edges and eventually (iii) outside its band. We obtain 
a very large enhancement of the typical thermopower at the band edges, while the sample to sample fluctuations around the 
typical value exhibit a sharp crossover from a Lorentzian distribution inside the impurity band towards a Gaussian distribution 
as the band edges are approached.
\end{abstract}
  
 
\maketitle

% ******************************* INTRO *********************************
\section{Introduction}
\begin{figure}
  \centering
  \includegraphics[keepaspectratio,width=0.8\columnwidth]{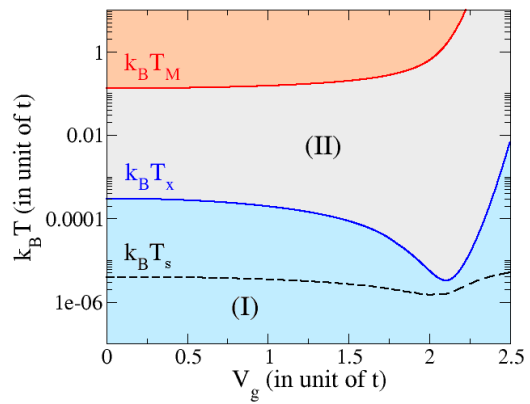}
  \caption{\label{fig_Tscale} 
(Color online) For a Fermi energy taken at the band center ($E_F=0$), the different regimes of electronic transport are given as a 
function of a positive gate voltage $V_g$. From bottom to top, one can see the elastic regime ($T<T_x$, blue), the inelastic VRH regime 
($T_x<T<T_M$, gray) and the simply activated regime ($T>T_M$, red)). The temperature scales $T_s$, $T_x=\xi/(2\nu N^2)$ and $T_M=2/(\xi\nu)$ 
are plotted for the 1D model introduced in Sec.~\ref{section_model} with $E_F=0$, $W=t$ and $N=1000$. $T_s$ is given for $\epsilon=0.01\%$ 
(see Sec.~\ref{section_Tc}). Transport exhibits the bulk behavior of the nanowire impurity band as far as $V_g$ does not exceed a value of 
order $1.5 t$ and its edge behavior in the interval $1.5t < V_g < 2.5t$. When $V_g> 1.5 t$, the bulk weak-disorder expansions (see 
section~\ref{section_model}) cease to be valid for $W=t$, while $ V_g > 2t+W/2=2.5t$ is necessary for completely depleting the nanowire 
in the limit $N \to \infty$. This paper is restricted to the study of region (I), corresponding to low
temperatures $T<T_s$ at which the Sommerfeld expansion can be applied for the calculation of the thermoelectric coefficients. The VRH region~(II) will be studied in Ref.~\cite{Bosisio2013}.
}
\end{figure}

Semiconductor nanowires emerged a few years ago as promising thermoelectric devices~\cite{Hicks1993}. In comparison to their bulk counterparts, 
they provide opportunities to enhance the dimensionless figure of merit $ZT=S^2\sigma T/\kappa$, which governs the efficiency of thermoelectric 
conversion at a given temperature $T$. Indeed, they allow one to reduce the phonon contribution $\kappa_{ph}$ to thermal conductivity 
$\kappa$~\cite{Hochbaum2008,Boukai2008,Martin2009}. On the other hand, through 
their highly peaked density of states they offer the large electron-hole asymmetry required for the enhancement of the thermopower $S$~\cite{Mahan1996,Tian2012}. 
This makes them now rank, with other nanostructured materials, among the best thermoelectrics in terms of achievable values of $ZT$. Yet, maximizing 
the figure of merit is not the ultimate requirement on the quest for improved thermoelectrics. The actual electric power that can be extracted from a 
heat engine (or conversely the actual cooling power that can be obtained from a Peltier refrigerator) is also of importance when thinking of 
practical applications. From that point of view, nanowire-based thermoelectric devices are also promising: they offer the scalability needed for 
increasing the output power, insofar as they can be arranged in arrays of nanowires in parallel.\\
\indent The main issue of this and the subsequent paper~\cite{Bosisio2013} is the determination of the dopant density optimizing the 
thermopower in a single semiconductor nanowire. From the theory side, this question has mainly been discussed at room temperature 
%thermoelectric conversion in a single semiconductor nanowire. From the theory side, this question has been mainly discussed at room temperature 
when the semi-classical Boltzmann 
theory can be used~\cite{Lin2000,Mingo2004,Neophytou2011} or in the ballistic regime~\cite{Liang2010} when the presence of the disorder is completely 
neglected. The goal was to describe the thermoelectric properties of nanowires at room temperature where the quantum effects become negligible, and 
in particular to probe the role of their geometry (diameter, aspect ratio, orientation, ...). From the experimental side, investigations have been 
carried out by varying the carrier density in the nanowire with an external gate electrode~\cite{Liang2009,Zuev2012,Tian2012,Moon2013,Wu2013,Roddaro2013}. 
Different field effect transistor device configurations can be used: either the nanowire and its metallic contacts are deposited on one side of 
an insulating layer, while a metallic back-gate is put on the other side (see for instance Refs.~\cite{Moon2013,Brovman2013}), or one can take  
a top-gate covering only the nanowire (see for instance Ref.~\cite{Poirier1999}). 
Recently, Brovman \textit{et al} have measured at room temperature the thermopower of Silicon and Silicon-Germanium nanowires and 
observed a strong increase when the nanowires become almost depleted under the application of a gate voltage~\cite{Brovman2013}. Interestingly, this work 
points out the importance of understanding thermoelectric transport near the band edges of semiconductor nanowires. It also reveals a lack of 
theoretical framework to this field that we aim at filling.\\
\indent In that purpose, we shall first identify as a function of the temperature $T$ and the applied gate voltage $V_g$ the dominant mechanism 
of electronic transport through a given nanowire. At low temperature $T<T_x$, transport is dominated by elastic tunneling processes and quantum 
effects must be properly handled. Due to the intrinsic disorder characterizing doped semiconductors, the electronic transport is much affected 
by Anderson localization while electron-phonon coupling can be neglected inside the nanowire. Above the activation temperature
$T_x$, electron-phonon coupling inside the nanowire start to be relevant. One enters the inelastic Variable Range Hopping (VRH) regime~\cite{Mott1979} 
where phonons help electrons to jump from one localized state to another, far away in space but quite close in energy. At temperatures higher than 
the Mott temperature $T_M$, the VRH regime ceases and one has simple thermal activation between nearest neighbor localized states. 
The different regimes are sketched in Fig.~\ref{fig_Tscale} for a nanowire modeled by a one-dimensional (1D) tight-binding 
Anderson model. Note that they are highly dependent on the gate voltage $V_g$. The inelastic VRH regime will be addressed in a subsequent paper~\cite{Bosisio2013}.\\
\indent In this work, we focus our study to the low temperature elastic regime or more precisely, to a subregion $T<T_s$ inside the elastic regime 
in which the thermopower can be evaluated using the Landauer-B\"uttiker scattering formalism and Sommerfeld expansions. An experimental study of 
the gate dependence of the electrical conductance of Si-doped GaAs nanowire in this elastic coherent regime can be found in Ref.\cite{Poirier1999}.\\
\indent We will mainly consider nanowires of size $N$ larger than their localization length $\xi$, characterized by exponentially small values of the electrical conductance. Obviously, this drastically reduces the output power associated with the thermoelectric conversion. 
Nevertheless, the advantage of considering the limit $N \gg \xi$ is twofold: first, the typical transmission at an energy $E$ is simply given by $\exp [-2N/\xi]$ in this limit, and second, at weak disorder, $\xi(E)$ is analytically known. This makes possible to derive analytical expressions describing the typical behavior of the thermopower.
To avoid the exponential reduction of the conductance at large $N/\xi$, one should take shorter lengths ($N \approx \xi$).  To study thermoelectric conversion in this crossover regime would require to use the scaling theory discussed in Ref.~\cite{Anderson1980, Pichard1986}. 
Furthermore, another reason to consider $N \gg \xi$ is that the delay time distribution (which probes how the scattering matrix depends on 
energy) has been shown to have a universal form~\cite{Texier1999} in this limit. We expect that this should be also the case for the 
fluctuations of the thermopower (which probes how the transmission depends on energy). This gives the theoretical reasons for focusing 
our study to the limit $N \gg \xi$.\\
\indent The outline of the manuscript is as follows. Section~\ref{section_LB} is a reminder about the Landauer-B\"uttiker formalism which allows one
to calculate thermoelectric coefficients in the coherent regime. In section~\ref{section_model}, we introduce the model and outline the numerical 
method used in this work, which is based on a standard recursive Green's function algorithm. Our results are presented in 
sections~\ref{section_Styp}, \ref{section_distrib} and~\ref{section_Tc}. Section~\ref{section_Styp} is devoted to the study of 
the typical behavior of the thermopower as the carrier density in the nanowire is modified with the gate voltage. We show that the thermopower 
is drastically enhanced when the nanowire is being depleted and we provide an analytical description 
of this behavior in the localized limit. In section~\ref{section_distrib}, we extend the study to the distribution of the thermopower. We show 
that the thermopower is always Lorentzian 
distributed, as long as the nanowire is not completely depleted by the applied gate voltage and provided it is long enough with respect to the 
localization length. Interestingly, the mesoscopic fluctuations appear to be basically larger and larger as the carrier density in the nanowire 
is lowered and the typical thermopower increases. As a matter of course, this ceases to be true when the gate voltage is so large that the 
nanowire, almost emptied of carriers, behaves eventually as a (disordered) tunnel barrier. In that case, the thermopower distribution is found 
to be Gaussian with tiny fluctuations. The evaluation of the ``crossover temperature'' $T_s$ (see Fig.~\ref{fig_Tscale}) is the subject of 
section~\ref{section_Tc}. Finally, we draw our conclusions in section~\ref{section_ccl}.

% ******************************* THERMOELECTRIC COEFFICIENTS *********************************

\section{Thermoelectric transport coefficients in the Landauer-B\"uttiker formalism}
\label{section_LB}
We consider a conductor connected via reflectionless leads to two reservoirs $L$ (left) and $R$ (right) in equilibrium at temperatures 
$T_L$ and $T_R$, and chemical potentials $\mu_L$ and $\mu_R$. To describe the thermoelectric transport across the conductor, we use the 
Landauer-B\"uttiker formalism~\cite{Datta1995}. The heat and charge transport are supposed to be mediated only by electrons and the phase 
coherence of electrons during their propagation through the conductor is supposed to be preserved. In this approach, the dissipation of 
energy takes place exclusively in the reservoirs while the electronic transport across the conductor remains fully elastic.
The method is valid as long as the phase-breaking length (mainly associated to electron-electron and electron-phonon 
interactions) exceeds the sample size. From a theoretical point of view, it can be applied to (effective) non-interacting models. In this 
framework, the electric ($I_e$) and heat ($I_Q$) currents flowing through the system are given by~\cite{Sivan1986,Butcher1990}
\begin{align}
I_e&=\frac{e}{h}\int\! dE \,\mathcal{T}(E)[f_L(E)-f_R(E)] \label{eq_Igeneral}\\
I_Q&=\frac{1}{h}\int\! dE \,(E-\mu_L)\mathcal{T}(E)[f_L(E)-f_R(E)] \label{eq_Jgeneral}
\end{align}
where $f_\alpha(E)=(1+\exp[(E-\mu_\alpha)/(k_B T_\alpha)])^{-1}$ is the Fermi distribution of the lead $\alpha$ and $\mathcal{T}(E)$ is the 
transmission probability for an electron to tunnel from the left to the right terminal. $k_B$ is the Boltzmann constant, $e<0$ the electron 
charge and $h$ the Planck constant. The above expressions are given for spinless electrons and shall be doubled in case of spin degeneracy.\\
\indent We now assume that the differences $\Delta\mu=\mu_L-\mu_R$ and $\Delta T=T_L-T_R$ to the equilibrium values $E_F\approx\mu_L\approx\mu_R$ 
and $T\approx T_L\approx T_R$ are small. Expanding the currents in Eqs.~(\ref{eq_Igeneral},\,\ref{eq_Jgeneral}) to first order in $\Delta\mu$ and 
$\Delta T$ around $E_F$ and $T$, one obtains~\cite{Butcher1990}
\be
\begin{pmatrix}
I_e \\
I_Q
\end{pmatrix} =
\begin{pmatrix}
L_0 & L_1 \\
L_1 & L_2
\end{pmatrix}
\begin{pmatrix}
\Delta \mu/eT \\
\Delta T/T^2
\end{pmatrix}
\ee
where the linear response coefficients $L_i$ are given by
\be
\label{eq_coeffLi}
L_i=\frac{e^2}{h}T\int\! dE \,\mathcal{T}(E)\left(\frac{E-E_F}{e}\right)^i\left(-\frac{\partial f}{\partial E}\right)\,.
\ee
The electrical conductance $G$, the electronic contribution $K_e$ to the thermal conductance $K$, the Seebeck coefficient $\mathcal{S}$ 
(or thermopower) and the Peltier coefficient $\Pi$ can all be expressed in terms of the Onsager coefficients $L_i$ as
\begin{align}
G&\equiv\left.\frac{eI_e}{\Delta \mu}\right|_{\Delta T=0}=\frac{L_0}{T}\label{eq_dfG}\\
K_e&\equiv\left.\frac{I_Q}{\Delta T}\right|_{I_e=0}=\frac{L_0L_2-L_1^2}{T^2L_0}\label{eq_dfkappa}\\
\mathcal{S}&\equiv-\left.\frac{\Delta \mu}{e\Delta T}\right|_{I_e=0}=\frac{L_1}{TL_0}\label{eq_dfS}\\
\Pi&\equiv\left.\frac{I_Q}{I_e}\right|_{\Delta T=0}=\frac{L_1}{L_0}~~.\label{eq_dfPi}
\end{align}
The Seebeck and Peltier coefficients turn out to be related by the Kelvin-Onsager relation~\cite{Onsager1931,Casimir1945}
\be
\label{eq_relPiS}
\Pi=\mathcal{S}T
\ee
as a consequence of the symmetry of the Onsager matrix. Note that, by virtue of Eq.~\eqref{eq_coeffLi}, in presence of particle-hole 
symmetry we have $\mathcal{S}=\Pi=0$. Further, the link between the electrical and thermal conductances is quantified by the Lorenz 
number $\mathcal{L}=K_e/GT$.\\
\indent In the zero temperature limit $T\to 0$, the Sommerfeld expansion~\cite{Ashcroft1976} can be used to estimate the 
integrals~\eqref{eq_coeffLi}. To the lowest order in $k_BT/E_F$, the electrical conductance reduces  
to $G\approx\frac{e^2}{h}\mathcal{T}(E_F)$ (ignoring spin degeneracy) while the thermopower simplifies to
\be
\label{eq_SeebeckMott}
\mathcal{S}\approx\frac{\pi^2}{3}\frac{k_B}{e}\,k_BT\,\left.\frac{\mathrm{d}\ln\mathcal{T}}{\mathrm{d}E}\right|_{E_F}\,.
\ee
The Lorenz number $\mathcal{L}$ takes in this limit a constant value,
\be
\label{eq_WFlaw}
\mathcal{L}\approx\mathcal{L}_0\equiv\frac{\pi^2}{3}\left(\frac{k_B}{e}\right)^2,
\ee
as long as $|\mathcal{S}|\ll\sqrt{\mathcal{L}_0}\simeq 156\,\mathrm{\mu V.K^{-1}}$. This reflects the fact that the electrical and thermal 
conductances are proportional and hence cannot be manipulated independently, an important although constraining property known as the 
Wiedemann-Franz (WF) law. This law is known to be valid for non-interacting systems if the low temperature Sommerfeld expansion is 
valid~\cite{Balachandran2012,Vavilov2005}, when Fermi liquid (FL) theory holds~\cite{Ashcroft1976,Chester1961} and for metals at room 
temperatures~\cite{Ashcroft1976}, while it could be largely violated in interacting systems due to non FL behaviors~\cite{Kane1996,Wakeham2011}.

% ******************************* MODEL AND METHOD *********************************

\section{Model and method}
\label{section_model}
\begin{figure}
  \centering
  \includegraphics[keepaspectratio,width=0.75\columnwidth]{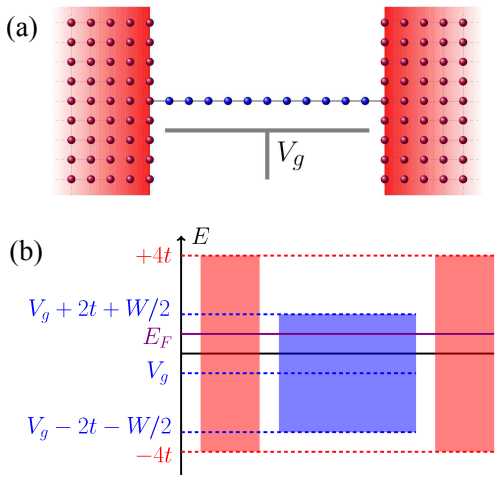}
  \caption{\label{fig_model} 
(Color online) (a) Sketch of the system: a 1D nanowire made of $N$ sites is connected to two leads at its extremities. An external gate voltage 
$V_g$ is applied. (b) Band diagram. The impurity band of the nanowire (in blue) can be shifted by the application of $V_g$ in order to probe 
either the bulk, the edges or the outside of the impurity band at Fermi energy $E_F$. Here, the leads are bidimensional (conduction band of the 
leads in red) and hence, $E_F\in[-4t,4t]$.}
\end{figure}

The system under consideration is sketched in Fig.~\ref{fig_model}(a). It is made of a 1D disordered nanowire coupled via perfect 
leads to two reservoirs $L$ (left) and $R$ (right) of non-interacting electrons, in equilibrium at temperature $T_L=T+\Delta T$ [$T_R=T$] and 
chemical potential $\mu_L=E_F+\Delta\mu$ [$\mu_R=E_F$]. The nanowire is modeled as a 1D Anderson chain of $N$ sites, with lattice spacing $a=1$. 
Its Hamiltonian reads,
\be
\label{eq_modelAnderson1D}
\mathcal{H}=-t\sum_{i=1}^{N-1}\left(c_i^{\dagger}c_{i+1}+\text{h.c.}\right)+\sum_{i=1}^{N}\epsilon_i c_i^{\dagger}c_i\,,
\ee
where $c^{\dagger}_i$ and $c_i$ are the creation and annihilation operators of one electron on site $i$ and $t$ is the hopping energy. 
The disorder potentials $\epsilon_i$ are (uncorrelated) random numbers uniformly distributed in the interval $[-W/2,W/2]$. The two sites at 
the ends of the nanowire are connected with hopping term $t$ to the leads which can be 1D semi-infinite chains or 2D semi-infinite square 
lattices, with zero on-site potentials and the same hopping term $t$. The simpler case of the Wide Band 
Limit (WBL) approximation, where the energy dependence of the self-energies of the leads is neglected, is also considered. Finally, an extra 
term 
\be
\mathcal{H}_{gate}=\sum_i V_g c_i^{\dagger}c_i
\ee
is added in the Hamiltonian~\eqref{eq_modelAnderson1D} to mimic the presence of an external metallic gate. 
It allows to shift the whole impurity band of the nanowire.

\subsection{Recursive Green's function calculation\\ of the transport coefficients}
In the Green's function formalism, the transmission $\mathcal{T}(E)$ of the system at an energy $E$ is given by the Fisher-Lee 
formula~\cite{Datta1995}
\be
\label{eq_FisherLee}
\mathcal{T}(E)=\mathrm{Tr}[\Gamma_L(E)G(E)\Gamma_R(E)G^\dagger(E)]
\ee
in terms of the retarded single particle Green's function $G(E)=[E-\mathcal{H}-\Sigma_L-\Sigma_R]^{-1}$ and of the retarded self-energies 
$\Sigma_L$ and $\Sigma_R$ of the left and right leads.
The operators $\Gamma_\alpha=i(\Sigma_\alpha-\Sigma_\alpha^\dagger)$ describe the coupling between the conductor and the lead $\alpha=L$ or $R$.  
A standard recursive Green's function algorithm~\cite{Lassl2007} allows us to compute the transmission $\mathcal{T}(E)$. The logarithmic 
derivative $\mathrm{d}\ln\mathcal{T}/\mathrm{d}E$ can be calculated as well with the recursive procedure, without need for a discrete 
evaluation of the derivative. It yields the thermopower $\mathcal{S}$ in the Mott-Sommerfeld approximation~\eqref{eq_SeebeckMott}. Hereafter, we will refer to a dimensionless thermopower
\be
\label{eq_df_S}
S=-t\left.\frac{\mathrm{d}\ln\mathcal{T}}{\mathrm{d}E}\right|_{E_F}
\ee
which is related, in the Mott-Sommerfeld approximation, to the true thermopower $\mathcal{S}$ as 
\be
\mathcal{S}=\frac{\pi^2}{3}\left(\frac{k_B}{|e|}\right)\left(\frac{k_BT}{t}\right)S\,.
\ee
We now discuss the expressions of the self-energies $\Sigma_L(E)$ and $\Sigma_R(E)$ of the left and right leads which are to be given as 
input parameters in the recursive Green's function algorithm. The nanowire of length $N$ sites is supposed to be connected on one site at 
its extremities to two identical leads, which are taken 1D, 2D or in the WBL approximation. Hence, the self-energies $\Sigma_\alpha$ (as well 
as the operator $\Gamma_\alpha$) are $N\times N$ matrices with only one non-zero component (identical for both leads) that we denote with 
$\Sigma$ (or $\Gamma$). When the wide-band limit is assumed for the leads, $\Sigma$ is taken equal to a small constant imaginary number 
independent of the energy $E$. When the leads are two 1D semi-infinite chains or two 2D semi-infinite square lattices, $\Sigma$ is given by 
the retarded Green's function $G_\mathrm{lead}$ of the lead under consideration evaluated at the site $X$ (in the lead) coupled to the nanowire, 
$\Sigma=t^2\langle X | G_\mathrm{lead} | X \rangle$. Knowing the expressions of the retarded Green's functions of the infinite 1D chain and the 
infinite 2D square lattice~\cite{Economou2006}, it is easy to deduce $G_\mathrm{lead}$ for the semi-infinite counterparts by using the method of 
mirror images. For 1D leads, one finds $\Sigma(E)=-te^{ik(E)}$ where $E=-2t\cos k$ and $k$ is the electron wavevector~\cite{Datta1995}. For 2D 
leads, the expression of $\Sigma(E)$ is more complicated (see Appendix~\ref{app_SelfNRJ}). As far as the Fermi energy $E_F$ is not taken near the 
edges of the conduction band of the leads, the thermopower behaviors using 1D and 2D leads coincide with those obtained using the WBL approximation 
(see Sec.~\ref{section_Styp}). This shows us that the dimensionality D becomes irrelevant in that limit, and we expect that taking 
3D leads will not change the results. 

\subsection{Scanning the impurity band of the Anderson model}
\label{subsec_dos}
\begin{figure}
  \centering
  \includegraphics[keepaspectratio,width=0.8\columnwidth]{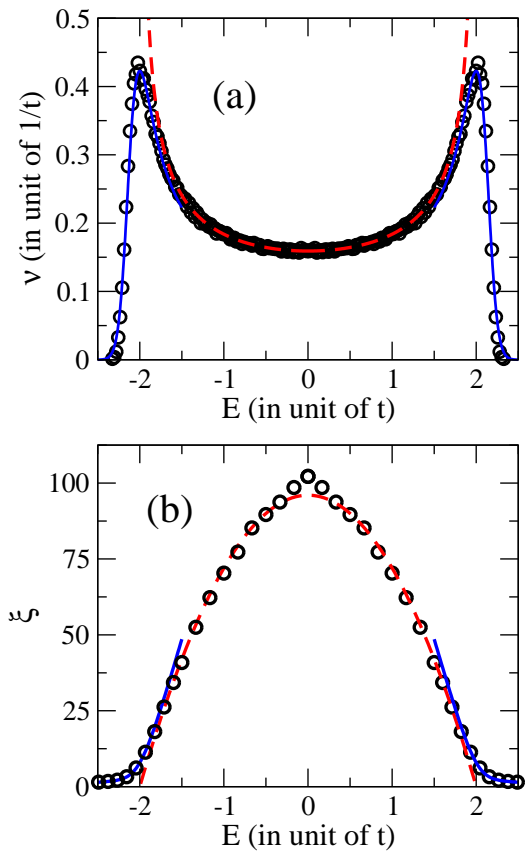}
  \caption{\label{fig_model2}
(a)  Density of states per site $\nu$ as a function of energy $E$ for the 1D Anderson model~\eqref{eq_modelAnderson1D} with disorder amplitude $W/t=1$. 
The circles correspond to numerical data (obtained with $N=1600$). The red dashed line and the blue line are the theoretical 
predictions~\eqref{eq_dstOfStateBulk} and~\eqref{eq_dstOfStateEdge}, expected in the bulk and at the edges of the nanowire conduction band 
for $N \to \infty$. (b) Localization length $\xi$ of the 1D Anderson model~\eqref{eq_modelAnderson1D} (with $W/t=1$) as a function of energy $E$. The circles 
correspond to numerical data (obtained with Eq.~\eqref{eq_typtrasm}). The red dashed line and the blue line are the theoretical 
predictions~\eqref{eq_xsi_bulk} and~\eqref{eq_xsi_edge} obtained in the limit $N \to \infty$.}
\end{figure}

The density of states per site $\nu(E)$ of the Anderson model, obtained by numerical diagonalization of the Hamiltonian~\eqref{eq_modelAnderson1D}, 
is plotted in Fig.~\ref{fig_model2}(a) in the limit $N \to \infty$. It is non-zero in the interval $[E_c^{-},E_c^{+}]$ where $E_c^{\pm}=\pm(2t+W/2)$ 
are the edges of the impurity band. In the bulk of the impurity band (\textit{i.e.} for energies $|E|\lesssim 1.5t$), the density of states is given 
with a good precision by the formula derived for a clean 1D chain (red dashed line in Fig.~\ref{fig_model2}(a)),
\be
\label{eq_dstOfStateBulk}
\nu_b(E)=\frac{1}{2\pi t\sqrt{1-(E/2t)^2}}~.
\ee
As one approaches the edges $E_c^{\pm}$, the disorder effect cannot be neglected anymore. The density of states is then well 
described by the analytical formula obtained by Derrida and Gardner around $E_c^{\pm}$, in the limit of weak disorder and large $N$ 
(see Ref.~\cite{Derrida1984}), 
\be
\label{eq_dstOfStateEdge}
\nu_e(E)=\sqrt{\frac{2}{\pi}}\left(\frac{12}{tW^2}\right)^{1/3}\frac{\mathcal{I}_1(X)}{[\mathcal{I}_{-1}(X)]^2}
\ee
where 
\be
X=(|E|-2t)t^{1/3}(12/W^2)^{2/3} 
\label{eq_scaling-variable}
\ee
and
\be
\label{eq_integralIn}
\mathcal{I}_n(X)=\int_0^{\infty} y^{n/2}\,e^{-\frac{1}{6}y^3+2Xy}\,dy\,.
\ee
\indent In this paper, we study the behavior of the thermoelectric coefficients as one probes at the Fermi energy $E_F$ electron transport either 
inside or outside the nanowire impurity band, and more particularly in the vicinity of its band edges. Such a scan of the impurity band can be 
done in two ways. One possibility is to vary the position of the Fermi energy $E_F$ in the leads. Doing so, we modify the distance between $E_F$ 
and the band edges $E_c^{\pm}$ but also the one between $E_F$ and the band edges of the leads. This can complicate the analysis of the data, the 
dimensionality of the leads becoming relevant when $|E_c^{\pm}-E_F|\to 0$. To avoid this complication, we can keep $E_F$ fixed far from 
$E_c^{\pm}$ and vary the gate voltage $V_g$ (see Fig.~\ref{fig_model}(b)).  

\subsection{Localization length of the Anderson model}
In the disordered 1D model~\eqref{eq_modelAnderson1D} we consider, all eigenstates are exponentially localized, with a localization length $\xi$. 
As a consequence, the typical transmission of the nanowire drops off exponentially with its length $N$. More precisely, when $N\gg\xi$ 
(localized limit), the distribution of $\ln\mathcal{T}$ is a Gaussian~\citep{Pichard1990,Pichard1991} centered around the value
\be\label{eq_typtrasm}
[\ln\mathcal{T}]_0(E)=-\frac{2N}{\xi(E)}\,,
\ee
as long as the energy $E$ of the incoming electron is inside the impurity band of the nanowire. The inverse localization length $1/\xi$ 
can be analytically obtained as a series of integer powers of $W$ when $W \to 0$. To the leading order (see e.g.~\cite{Kramer1993}), this 
gives 
\be
\label{eq_xsi_bulk}
\xi_b(E)\approx \frac{24}{W^2}\left(4t^2-E^2\right)\,.
\ee
The formula is known to be valid in the weak disorder limit inside the bulk of the impurity band (hence the index $b$). Strictly speaking, it fails 
in the vicinity of the band center $E=0$ where the perturbation theory does not converge~\cite{Kappus1981} but it gives nevertheless a good 
approximation. As one approaches one edge of the impurity band, the coefficients characterizing the expansion of $1/\xi$ in integer powers of $W$ 
diverge and the series has to be reordered. As shown by Derrida and Gardner~\cite{Derrida1984}, this gives (to leading order in $W$) the non 
analytical behavior $1/\xi \propto W^{2/3}$ as one edge is approached instead the analytical behavior $1/\xi \propto W^2$ valid in the bulk of 
the impurity band. More precisely, one find in the limit $W \to 0$ that
\be
\label{eq_xsi_edge}
\xi_e(E)=2\left(\frac{12t^2}{W^2}\right)^{1/3}\frac{\mathcal{I}_{-1}(X)}{\mathcal{I}_{1}(X)}
\ee
as $E$ approaches the band edges $\pm 2t$. The integrals $\mathcal{I}_i$ and the parameter $X$ have been defined in Eq.~\eqref{eq_integralIn} and 
Eq.~\eqref{eq_scaling-variable}. As shown in Fig.~\ref{fig_model2}(b), both formula~\eqref{eq_xsi_bulk} and~\eqref{eq_xsi_edge} are found to be in very 
good agreement with our numerical evaluation of $\xi(E)$, in the respective range of energy that they describe, even outside a strictly weak 
disorder limit ($W=t$ in Fig.~\ref{fig_model2}(b)).

% ******************************* TYPICAL THERMOPOWER *********************************

\section{Typical thermopower}
\label{section_Styp} 
\begin{figure}
  \centering
  \includegraphics[keepaspectratio,width=\columnwidth]{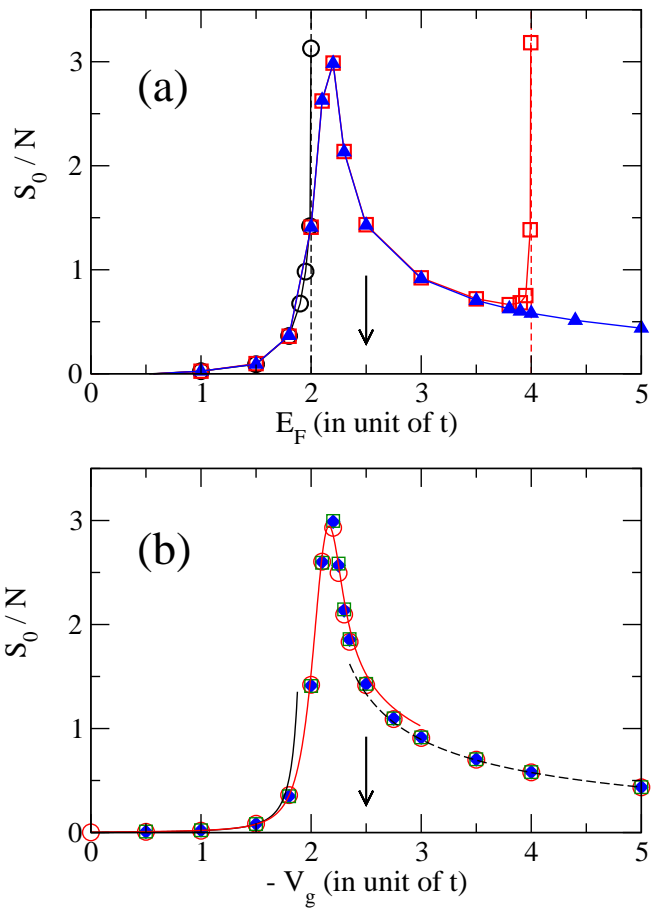}
  \caption{\label{fig_Styp} 
(Color online) Typical value of the dimensionless thermopower per unit length, $S_0/N$, as a function of the Fermi energy $E_F$ at $V_g=0$ (a) and 
as a function of the gate voltage $V_g$ at $E_F=0$ (b). In panel~(a), the data were obtained at fixed $N=500$, by using either 1D 
leads~({\large$\circ$}), 2D leads~({\tiny{\color{red}$\square$}}) or the wide-band limit approximation~({\scriptsize{\color{blue}$\blacktriangle$}}). 
With 1D [2D] leads, the typical thermopower shows a divergent behavior at the band edge of the leads (black [red] vertical dashed line). 
In panel~(b), 1D leads are used. The symbols stand for different lengths of the nanowire ($N=200$~({\large{\color{red}$\circ$}}), 
$800$~({\tiny{\color{DarkGreen}$\square$}}) and $1600$~({\scriptsize{\color{blue}$\blacklozenge$}})). The full black line, the full red line 
and the dashed black line correspond respectively to the theoretical fits~\eqref{eq_S0bulk},~\eqref{eq_S0edge} and~\eqref{eq_S0TB} expected when 
$E_F$ probes the bulk, the edge and the outside of the impurity band. In both panels, $W/t=1$. The arrows indicate the position of the edge of the 
impurity band of the nanowire.}
\end{figure}

We compute numerically the thermopower $S$ for many realizations of the disorder potentials $\epsilon_i$ in Eq.~\eqref{eq_modelAnderson1D}, and we 
define the \emph{typical} value $S_0$ as the median of the resulting distribution $P(S)$. As it will be shown in Sec.~\ref{section_distrib}, 
$P(S)$ is typically a smooth symmetric function (Lorentzian or Gaussian), and thus its median coincides with its most probable value. We study 
the behavior of $S_0$ as one scans the energy spectrum of the nanowire by varying the position of the Fermi energy $E_F$ in the leads or the 
gate voltage $V_g$.\\
\indent In Fig.~\ref{fig_Styp}(a), the typical thermopower $S_0$ of a long nanowire in the localized regime ($N \gg \xi$) is plotted as a function 
of $E_F$ without gate voltage ($V_g=0$). Since $S_0\to -S_0$ when $E_F\to -E_F$, data are shown for positive values of $E_F$ only. In the figure, 
three different kinds of leads are considered: 1D leads, 2D leads or leads in the WBL approximation. In all cases, as expected, we find that $S_0=0$ 
at the center of the conduction band of the leads ($E_F=0$). Indeed, the random potentials being 
symmetrically distributed around a zero value, one has a statistical particle-hole symmetry at the band center and the thermopower can only be a statistical 
fluctuation around a zero typical value. As $E_F$ is increased, the statistical particle-hole symmetry breaks down and $S_0$ gets finite. Here 
$S_0>0$ because charge transport is dominated by holes for $E_F>0$. When the wide band limit is assumed for both leads (triangles in 
Fig.~\ref{fig_Styp}(a)), we find that the typical thermopower $S_0$ increases with $E_F$ and reaches a maximum just before $E_c^+=2t+W/2$, the asymptotic 
$N\to\infty$ value for the edge ($E_c^+=2.5\,t$ in Fig.~\ref{fig_Styp}(a) where $W=t$) before decreasing. The same curve is obtained with 1D [2D] 
leads as long as the Fermi energy $E_F$ remains far enough below the upper band edge of the $D$-dimensional leads. When $E_F$ approaches $2t$ [$4t$], the 
typical thermopower $S_0$ of the nanowire is found to increase drastically, contrary to the WBL case (of course, no data are available for 
$|E_F|\geq 2t\,[4t]$, charge transfer being impossible outside the conduction band of the leads). This singularity at the band edge of the leads 
can be easily understood using Eqs.~\eqref{eq_FisherLee} and~\eqref{eq_df_S} and noticing that for 1D [2D] leads, 
$\mathrm{d}\ln\Gamma/\mathrm{d}E\to -\infty$ as 
$E\to 2t~[4t]$. This is obvious in the case of 1D leads where $\Gamma(E)=2t\sqrt{1-(E/2t)^2}$ and it can also be shown for 2D leads. We will see 
in Sec.~\ref{section_Tc} that this apparent divergence of the thermopower is actually only valid in an infinitesimally small range of temperatures 
above $0\,$K.\\
\indent With the gate voltage $V_g$, we can explore the impurity band of the nanowire while keeping $E_F$ fixed. The behavior of $S_0$ as a function 
of $V_g$ is shown in Fig.~\ref{fig_Styp}(b) for $E_F=0$ and 1D leads. It is found to be identical to the behavior of $S_0$ as a function of $E_F$ 
obtained at $V_g=0$ in the WBL approximation. This remains true if 2D leads are used in Fig.~\ref{fig_Styp}(b) and we have no doubt that it also 
remains true with 3D leads. Moreover, the results are unchanged if $E_F$ is fixed to any other value, as long as it does not approach too closely 
one edge of the conduction band of the leads (but it can be chosen close enough to one band edge to recover the continuum limit of the leads). Our 
main observation is that the typical thermopower $S_0$ increases importantly when the Fermi energy probes the region around the edges of the impurity 
band of the nanowire. Qualitatively, this is due to the fact that the typical transmission of the nanowire drops down when the edges are 
approached: this huge decrease results in a enhancement of the typical thermopower, the thermopower being somehow a measure of the energy dependence 
of the transmission. A quantitative description of this behavior can also be obtained. Indeed, since the distribution of the transmission 
$\mathcal{T}$ is log-normal in the localized regime~\citep{Pichard1990,Pichard1991} and the thermopower $S$ is calculated for each disorder configuration 
with the Mott approximation~\eqref{eq_df_S}, one expects to have  
\be
S_0=-t\left.\frac{\mathrm{d}[\ln\mathcal{T}]_0}{\mathrm{d}E}\right|_{E_F}
\ee
where $[\ln\mathcal{T}]_0$ is the median of the $\ln\mathcal{T}$ Gaussian distribution (which in this case coincides with the most probable value). 
Moreover, according to Eq.~\eqref{eq_typtrasm}, the energy dependence of $[\ln\mathcal{T}]_0$ is given by the 
energy dependence of the localization length, \textit{i.e.} by Eqs.~\eqref{eq_xsi_bulk} and~\eqref{eq_xsi_edge}. This allows us to derive the 
following expressions for the typical thermopower in the bulk and at the edges:
\be
\label{eq_S0bulk}
S_0^b=N\frac{(E_F-V_g)\,W^2}{96t^3[1-((E_F-V_g)/2t)^2]^2},
\ee
\be
\label{eq_S0edge}
S_0^e=2N\left(\frac{12t^2}{W^2}\right)^{1/3}\left\{ \frac{\mathcal{I}_{3}(X)}{\mathcal{I}_{-1}(X)}-\left[\frac{\mathcal{I}_{1}(X)}{\mathcal{I}_{-1}(X)}\right]^2\right\},
\ee
where now $X$ is modified to 
\be
X=(|E_F-V_g|-2t)t^{1/3}(12/W^2)^{2/3}
\ee
in order to take into account the effect of the gate voltage $V_g$.
When the outside of the impurity band, rather than the inside, is probed at $E_F$ (i.e. when the wire is completely depleted), no more states 
are available in the nanowire to tunnel through. Electrons coming from one lead have to tunnel directly to the other lead through the disordered 
barrier of length $N$. We have also calculated the typical thermopower of the nanowire in that case, assuming that the disorder effect is 
negligible (see Appendix~\ref{appThermopowerCTB}). We find
\be
\label{eq_S0TB}
\frac{S_0^{TB}}{N} \underset{N\to\infty}{\approx} -\frac{1}{N}\frac{2t}{\Gamma(E_F)}\left.\frac{\mathrm{d}\Gamma}{\mathrm{d}E}\right|_{E_F}\mp\frac{1}{\sqrt{\left(\frac{E_F-V_g}{2t}\right)^2-1}}
%\left[\left(\frac{E_F-V_g}{2t}\right)^2-1\right]^{-1/2}
\ee
with a $+$ sign when $E_F\leq V_g-2t$ and a $-$ sign when $E_F\geq V_g+2t$. Fig.~\ref{fig_Styp}(b) shows a very good agreement between the 
numerical results (symbols) and the expected behaviors (Eqs.~\eqref{eq_S0bulk},~\eqref{eq_S0edge} and~\eqref{eq_S0TB}). One consequence of these 
analytical predictions is that the peak in the thermopower curves gets higher and narrower as the disorder amplitude is decreased (and vice-versa).

%We note incidentally that our findings look in qualitative agreement with the recent experimental observation reported in Ref~\cite{Brovman2013}. 
%We stress out however that those measurements were carried out outside the low temperature coherent regime which consider, at room temperatures. 
%To describe them, inelastic effects must be included. It will be the purpose of our next paper~\cite{Bosisio2013}. 

% ******************************* THERMOPOWER DISTRIBUTIONS *********************************

\section{Thermopower distributions}
\label{section_distrib}

\begin{figure}
  \centering
  \includegraphics[keepaspectratio, width=\columnwidth]{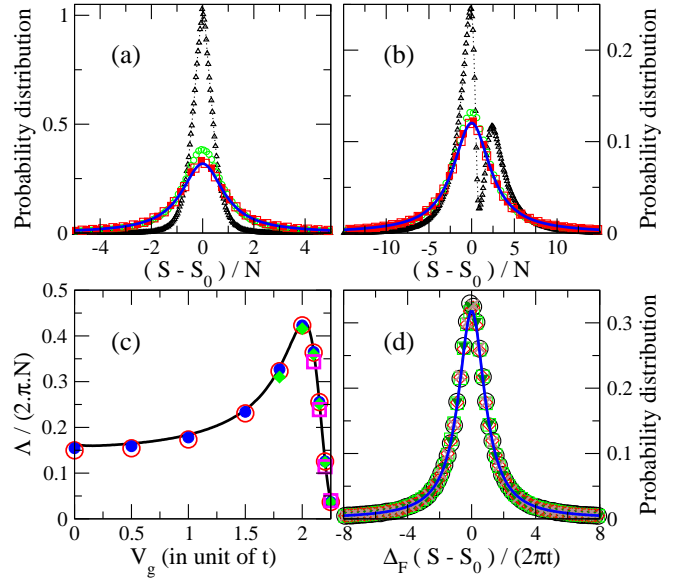}
  \caption{\label{fig_distribS_lor} 
(Color online) Top panels: probability distributions of the rescaled thermopower $(S-S_0)/N$ at $V_g=0$ (a) and $V_g=2t$ (b), with $W=t$, $E_F=0$ 
and 1D leads. In each panel, the different symbols correspond to various lengths of the chain ($N\approx\xi$~({\tiny{\color{black}$\triangle$}}), 
$N\approx 10\,\xi$~({\small{\color{green}$\circ$}}), $N\approx 50\,\xi$~({\scriptsize{\color{red}$\square$}}) and 
$N\approx 100\,\xi$~({\tiny{\color{red}$\blacksquare$}}), respectively $N=100$, $1000$, $5000$ and $10000$ in (a) and 
$N=10$, $100$, $500$ and $1000$ in (b). The distributions obtained for $N\geq 50\,\xi$ collapse on a single curve which is well fitted by a 
Lorentzian distribution function (thick blue lines). The widths $\Lambda/N$ of the Lorentzian fits are plotted as a function of $V_g$ in panel 
(c), for $N=200$ ({\scriptsize{\color{magenta}$\square$}}), $1000$ ({\scriptsize{\color{green}$\blacklozenge$}}), $5000$ 
({\Large{\color{red}$\circ$}}) and $10000$ ({\large{\color{blue}$\bullet$}}), together with the density of states per site at $E_F$, 
$t\nu_F$, of the closed chain (red line). The probability distributions of the rescaled thermopower $(\Delta_F/2\pi t)(S-S_0)$, obtained in the 
large $N$ limit ($N\approx 100\,\xi$) and for various sets of parameters ($W=0.5t$ and $V_g=2t$~({\large{\color{red}$\diamond$}}), $W=t$ and 
$V_g=0$~({\Large{\color{black}$\circ$}}), $W=t$ and $V_g=2t$~({\scriptsize{\color{green}$\square$}}), $W=2t$ and 
$V_g=0$~({\small{\color{brown}$\times$}}), and $W=2t$ and $V_g=2.3t$~({\scriptsize{\color{DarkGreen}$\blacktriangledown$}}), with 
$E_F=0$ in all cases), are shown in panel (d). They all collapse on the blue line which is the Lorentzian function $y=1/[\pi(1+x^2)]$. }
\end{figure}

In the coherent elastic regime we consider, the sample-to-sample fluctuations of the thermopower around its typical value are expected to be large. 
The most striking illustration occurs at the center of the impurity band of the nanowire ($E_F=V_g$), when the typical thermopower is zero due to 
statistical particle-hole symmetry but the mesoscopic fluctuations allow for large thermopower anyway. Van Langen \textit{et al} showed 
in Ref.~\cite{VanLangen1998} that in the localized regime $N\gg\xi$ without gate ($V_g=0$) and around the band center ($E_F\approx 0$), the 
distribution of the low-temperature thermopower is a Lorentzian,
\be
\label{eq_distrib_lor}
P(S)=\frac{1}{\pi}\frac{\Lambda}{\Lambda^2+(S-S_0)^2}\,,
\ee
with a center $S_0=0$ and a width
\be
\label{eq_width_lor}
\Lambda=\frac{2\pi t}{\Delta_F}
\ee
given by $\Delta_F=1/(N\nu_F)$, the average mean level spacing at $E_F$. This was derived under certain assumptions leading to $S_0=0$.  
As we have shown, $S_0=0$ is exact only at the impurity band center ($E_F=0$ when $V_g=0$) and remains a good approximation as far as 
one stays in the bulk of the impurity band. But the distribution $P(S)$ is no more centered around zero as one approaches the band edge.
\\
\indent We propose here to investigate how the thermopower distribution $P(S)$ is modified when this is not only the bulk, but the edges (or 
even the outside) of the impurity band which are probed at the Fermi energy $E_F$. To fix the ideas, we set the Fermi 
energy to $E_F=0$ and the disorder amplitude to $W=t$ (so that the band edges are $V_g+E_c^\pm=V_g\pm2.5t$). First, we check in 
Fig.~\ref{fig_distribS_lor}(a) that at $V_g=0$ and in the localized regime, the thermopower distribution is indeed a Lorentzian with a width 
$\Lambda\propto N$. We note that very long chains of length $N\approx 50\xi$ ($\xi\approx100$ here) are necessary to converge to the 
Lorentzian ~\eqref{eq_distrib_lor}. Moreover, we have checked that this is also in this limit that the delay time distribution converges towards 
the universal form predicted in Ref~\cite{Texier1999}.

\begin{figure}
  \centering
  \includegraphics[keepaspectratio, width=\columnwidth]{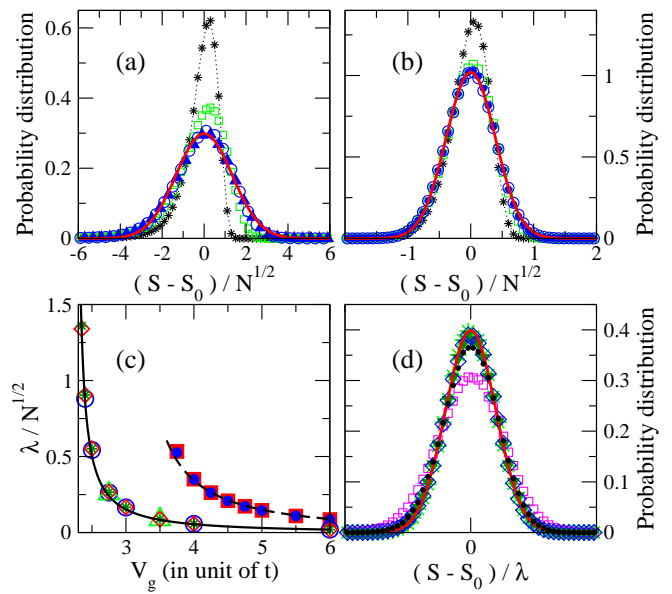}
  \caption{\label{fig_distribS_gauss} 
(Color online) Top panels: probability distributions of the rescaled thermopower $(S-S_0)/\sqrt{N}$ at $V_g=2.35t$ (a) and $V_g=2.6t$ (b), 
with $W=t$, $E_F=0$ and 1D leads. In each panel, the distributions are plotted for various lengths of the chain 
($N=10$~({\normalsize{\color{black}$\ast$}}), $50$~({\tiny{\color{green}$\square$}}), $200$~({\small{\color{blue}$\bullet$}}), 
$500$~({\normalsize{\color{blue}$\circ$}}) and $1000$~({\small{\color{blue}$\blacktriangle$}})) and collapse at large $N$ on one single curve, 
well fitted by a Gaussian distribution (red line). The widths $\lambda/\sqrt{N}$ of the Gaussian fits are plotted as a function of $V_g$ in 
panel (c), for various lengths ($N=50$ (triangle), $200$ (circle), $400$ (square), $800$ (diamond) and $1600$ (star)) and two disorder 
amplitudes ($W=t$ (empty symbols) and $W=4t$ (full symbols)). The solid and dashed lines are the fits given by Eq.~\eqref{eq_width_gauss}, 
respectively for $W=t$ and $W=4t$. Panel (d): collapse of the thermopower distributions, obtained with $N=500$ and various parameters 
($W=0.5t$ and $V_g=2.25t$ ({\tiny{\color{magenta}$\square$}}), $W=0.5t$ and $V_g=5t$ ({\small{\color{DarkGreen}$\blacktriangledown$}}), 
$W=t$ and $V_g=2.5t$ ({\small{\color{black}$\bullet$}}), $W=t$ and $V_g=5t$ ({\normalsize{\color{green}$\ast$}}), and $W=4t$ and 
$V_g=4t$ ({\scriptsize{\color{blue}$\lozenge$}})), after a rescaling by $\lambda$ as given in Eq.~\eqref{eq_width_gauss}. The red line is 
the Gaussian distribution $y=(1/\sqrt{2\pi})\exp(-x^2/2)$.}
\end{figure}

Then we increase the gate potential up to $V_g=2t$ to approach the edge $E_c^-$ of the impurity band and find that the thermopower distribution 
remains a Lorentzian in the localized regime ($N\gtrsim 50\xi$) with a width $\Lambda\propto N$, as shown in Fig.~\ref{fig_distribS_lor}(b). It 
turns out actually that the fit of the thermopower distribution with a lorentzian (in the large $N$ limit) is satisfactory in a broad range of 
gate potentials $|V_g|\lesssim 2.25t$, as long as the Fermi energy $E_F=0$ probes the impurity band without approaching too closely 
its edges $V_g+E_c^\pm$. In Fig.~\ref{fig_distribS_lor}(c), we show in addition that in this regime, the widths $\Lambda$ of the Lorentzian fits to 
the thermopower distributions $P(S)$ obey $\Lambda/(2\pi Nt)=\nu_F$, \textit{i.e.} Eq.~\eqref{eq_width_lor}. 
Therefore (Fig.~\ref{fig_distribS_lor}(d)), we can use this parameter to rescale all the distributions obtained in a broad range of parameters, 
on the same Lorentzian function $y=1/[\pi(1+x^2)]$. A direct consequence of~Eq.~\eqref{eq_width_lor} is that the mesoscopic fluctuations of the 
thermopower are maximal for $|E_F-V_g|\approx 2t$.\\
\indent When the gate voltage $|V_g|$ is increased further, the number of states available at $E_F$ in the nanowire decreases exponentially and 
eventually vanishes: one approaches eventually a regime where the nanowire becomes a long tunnel barrier and where 
the thermopower fluctuations are expected to be smaller and smaller. In this limit, we find that the thermopower distribution is no more a 
Lorentzian but becomes a Gaussian,
\be
\label{eq_distrib_gauss}
P(S)=\frac{1}{\sqrt{2\pi}\lambda}\exp\left[-\frac{(S-S_0)^2}{2\lambda^2}\right]\,,
\ee
provided the chain is long enough. This result is illustrated in Figs.~\ref{fig_distribS_gauss}(a) and~\ref{fig_distribS_gauss}(b) for 
two values of $V_g$. The Gaussian thermopower distribution is centered around a typical value $S_0$ given by Eq.~\eqref{eq_S0TB} and its 
width $\lambda$ is found with great precision to increase linearly with $\sqrt{N}$ and $W$. To be more precise, we find that the dependency of 
$\lambda$ on the various parameters is mainly captured by the following formula
\be
\label{eq_width_gauss}
\lambda\approx0.6\frac{Wt\sqrt{N}}{\left(E_F-V_g\right)^2-\left(2t+W/4\right)^2}\,,
\ee
at least for $0.5t\lesssim W \lesssim 4t$, $ 2.35t\lesssim |E_F-V_g|\lesssim 6t$ and $N\gtrsim 100$ (see Fig.~\ref{fig_distribS_gauss}(c)). 
We stress out that Eq.~\eqref{eq_width_gauss} is merely a compact way of describing our numerical data. In particular, the apparent divergence 
of $\lambda$ when $|E_F-V_g|\to 2t+W/4$ is meaningless and in fact, it occurs outside the range of validity of the fit. To double-check the 
validity of Eq.~\eqref{eq_width_gauss}, we have rescaled with the parameter $\lambda$ given by Eq.~\eqref{eq_width_gauss}, a set of thermopower 
distributions obtained in the disordered tunnel barrier regime, for various $W$ and $V_g$. All the resulting curves (plotted in 
Fig.~\ref{fig_distribS_gauss}(d)) are superimposed on the unit gaussian distribution, except the one for the smallest disorder value $W=0.5t$ for 
which the fit~\eqref{eq_width_gauss} to $\lambda$ is satisfactory but not perfect. 

\begin{figure}
  \centering
  \begin{psfrags}	
  \psfrag{XXX}{\footnotesize{$W$ (in unit of $t$)}}
  \psfrag{YYY}{\footnotesize{$V_g$ (in unit of $t$)}}
  \psfrag{Y}{\footnotesize{$\eta$}}
  \includegraphics[keepaspectratio,trim = 0mm 0mm 2mm 3mm, clip, width=\columnwidth]{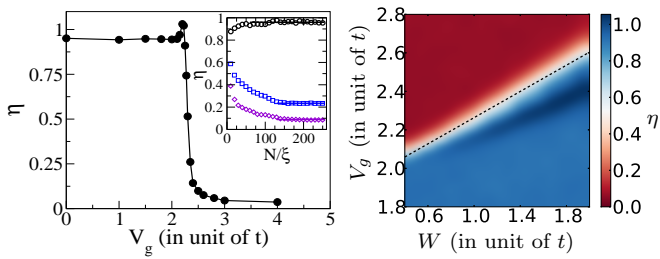}
  \end{psfrags}
  \caption{\label{fig_eta} 
(Color online) Left panel: in the inset, $\eta$ parameter as a function of $N/\xi$ for various gate voltages 
($V_g=1.9\,t$~($\circ$), $2.35\,t$~({\tiny{\color{blue}$\square$}}) and $2.5\,t$~({\color{violet}$\diamond$})), at $E_F=0$ and $W/t=1$. 
The horizontal lines show the convergence of $\eta$ at large $N$. The asymptotic values are plotted in the main panel as a function of $V_g$. 
Right panel: $\eta$ parameter in the limit of large $N$ as a function of $V_g$ and $W$, at $E_F=0$. Upon shifting the spectrum of the nanowire 
with $V_g$, the thermopower distribution moves from a Lorentzian distribution for $V_g\lesssim V_g^c$ ($\eta\approx 1$, blue) to a Gaussian 
distribution for $V_g\gtrsim V_g^c$ ($\eta\approx 0$, red), where $V_g^c=1.92t+0.34W$ (dashed line).}
\end{figure}

To identify precisely the position of the crossover between the Lorentzian regime and the Gaussian regime, we introduce now the parameter 
$\eta$,
\be
\label{eq_df_eta}
\eta=\frac{\int dS|P(S)-P_G(S)|}{\int dS|P_L(S)-P_G(S)|}\,,
\ee
which measures, for a given thermopower distribution $P(S)$ obtained numerically, how closed it is from its best Gaussian fit $P_G(S)$ and 
from its best Lorentzian fit $P_L(S)$\footnote{One could be tempted to compare an arbitrary thermopower distribution $P(S)$ to the Lorentzian 
and Gaussian distributions given in Eqs.~(\ref{eq_distrib_lor}\,-\,\ref{eq_width_lor}) and~(\ref{eq_distrib_gauss}\,-\,\ref{eq_width_gauss}) 
respectively. However, to define $\eta$ for any set of parameters, one should extend to the outside of the spectrum the 
formula~\eqref{eq_width_lor} for the width $\Lambda$ of the Lorentzian, and to the inside of the spectrum the formula~\eqref{eq_width_gauss} 
for the width $\lambda$ of the Gaussian. We avoid this problem by taking instead the best Lorentzian and Gaussian fits to $P(S)$ in the definition 
of $\eta$. It allows us to distinguish whether $P(S)$ is a Lorentzian or a Gaussian (or none of both) but of course, the precise form of $P(S)$ is 
not probed by $\eta$ as defined.}. If $P(S)$ is a Lorentzian, $\eta=1$ while $\eta=0$ if it is a Gaussian. Considering first the case where $E_F=0$ 
and $W=t$, we show in the left panel of Fig.~\ref{fig_eta} that $\eta$ converges at large $N$ for any $V_g$ (inset). The asymptotic values of 
$\eta$ (given with a precision of the order of $0.05$ in the main panel) undergo a transition from $\eta\approx 1$ to $\eta\approx 0$ when $V_g$ is 
increased from $0$ to $4t$. This reflects the crossover from the Lorentzian to the Gaussian thermopower distribution already observed in the top 
panels of Figs.~\ref{fig_distribS_lor} and~\ref{fig_distribS_gauss}. We see in addition that the crossover is very sharp around the value 
$V_g\approx 2.3t$, indicating a crossover which remains inside the impurity band of the infinite nanowire, since the band is not shifted enough when 
$V_g\approx 2.3t$ to make the Fermi energy coincides with the band edge $V_g+E_c^- = V_g-2.5t$. We have obtained the same results for other values of 
the disorder amplitude. After checking the convergence of $\eta$ at large $N$, we observe the same behavior of the asymptotic values of $\eta$ as 
a function of $V_g$, for any $W$. Only the position of the crossover is disorder-dependent. Those results are summarized in the right panel of 
Fig.~\ref{fig_eta} where one clearly sees the crossover (in white) between the Lorentzian regime (in blue) and the Gaussian regime (in red). 
It occurs around $V_g\approx 1.92t+0.34W$, not exactly when $E_F=V_g+E_c^-$, but in a region where the number of states available at $E_F$ in the 
nanowire becomes extremely small. To be precise, we point out that the values of $\eta$ in the 2D colorplot are given with a precision of the 
order of $0.1$. Hence, one cannot exclude that the white region corresponding to the crossover actually reduces into a single line $V_g^c(W)$. 
One could also conjecture the existence of a third kind of thermopower distribution (neither Lorentzian, nor Gaussian) associated to this critical 
value $V_g^c$. Our present numerical results do not allow to favor one scenario (sharp crossover) over the other (existence of a critical edge 
distribution).\\

% ******************************* TEMPERATURE SCALE *********************************

\section{Temperature range of validity of the Sommerfeld expansion}
\label{section_Tc}
\begin{figure}
  \centering
  \includegraphics[keepaspectratio, width=\columnwidth]{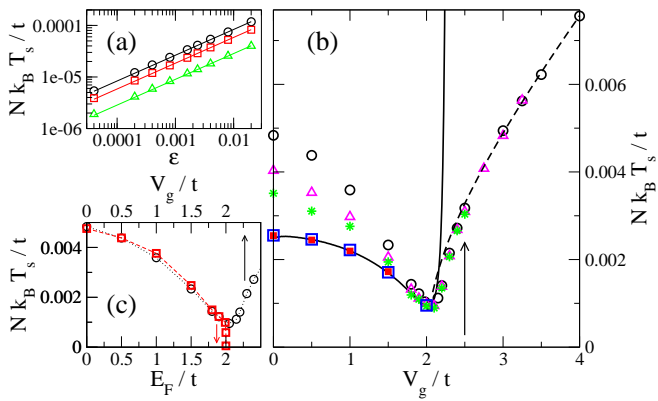}
  \caption{\label{fig_Tc} 
(Color online) Temperature scale $T_s$ above which the WF law breaks down. (a)~$Nk_BT_s/t$ as a function of the desired precision 
$\epsilon$ on $\mathcal{L}$. The critical temperatures were extracted for different values of $V_g$ ($V_g=t$~({\large{\color{black}$\circ$}}), 
$1.5t$~({\tiny{\color{red}$\square$}}), $2.02t$~({\tiny{\color{green}$\triangle$}})), with $E_F=0$, $W/t=1$, $N=500$ and 1D leads. The 
solid lines are fits $T_s=T_s^*\sqrt{\epsilon}$. (b)~$Nk_BT_s/t$ (extracted for $\epsilon=4\times 10^{-5}$) as a function of $V_g/t$, for 
chains of different length ($N=150$~({\large{\color{black}$\circ$}}), $300$~({\scriptsize{\color{magenta}$\triangle$}}), 
$500$~({\normalsize{\color{green}$\ast$}}), $1500$~({\small{\color{blue}$\square$}}) and $3000$~({\tiny{\color{red}$\blacksquare$}})), 
with $E_F=0$, $W/t=1$ and 1D leads. The solid line is $4.04\times 10^{-4}/(\nu_F t)$, the dashed line is $4.37\times 10^{-3}\sqrt{(V_g/2t)^2-1}$ 
and the arrow indicates the position of the edge of the impurity band. (c)~$Nk_BT_s/t$ (extracted for $\epsilon=4\times 10^{-5}$) as a function 
of $E_F/t$ at $V_g=0$ ({\scriptsize{\color{red}$\square$}}) and as a function $V_g/t$ at $E_F=0$ ({\large{\color{black}$\circ$}}), with $N=150$, 
$W/t=1$ and 1D leads. Dashed lines are guides to the eye.} 
\end{figure}

All the results discussed in this paper have been obtained in the low temperature limit, after expanding the thermoelectric coefficients to 
the lowest order in $k_BT/E_F$. To evaluate the temperature range of validity of this study, we have calculated the Lorenz number 
$\mathcal{L}=K_e/GT$ beyond the Sommerfeld expansion, and looked at its deviations from the WF law $\mathcal{L}=\mathcal{L}_0$ 
(see Eq.~\eqref{eq_WFlaw}): We have computed numerically the integrals~\eqref{eq_coeffLi} enterings Eqs.~\eqref{eq_dfG} and~\eqref{eq_dfkappa}, 
deduced $\mathcal{L}(T)$ for increasing values of temperature, and then recorded the temperature $T_s$ above which $\mathcal{L}(T)$ differs from 
$\mathcal{L}_0$ by a percentage $\epsilon$, $\mathcal{L}(T_s)=\mathcal{L}_0(1\pm\epsilon)$. We did it sample by sample and deduced the temperature 
$T_s$ averaged over disorder configurations. Our results are summarized in Fig.~\ref{fig_Tc}.\\
\indent In panel~(a), we analyze how sensitive $T_s$  is to the precision $\epsilon$ on the Lorenz number $\mathcal{L}$. We find that $T_s$ 
increases linearly with $\sqrt{\epsilon}$, $T_s(\epsilon)=T_s^*\sqrt{\epsilon}$, at least for $\epsilon\leq 2\%$. This is not surprising since 
the Sommerfeld expansion leads to $\mathcal{L}-\mathcal{L}_0\propto (k_BT)^2$, when one does not stop the expansion to the leading order in 
temperature ($\mathcal{L}=\mathcal{L}_0$) but to the next order.\\
\indent The main result of this section is shown in Fig.~\ref{fig_Tc}(b) where we have plotted the temperature $T_s$ as a function of the gate 
 voltage $V_g$, for chains of different lengths, at fixed $E_F=0$ and $W=t$. As long as the Fermi energy probes the inside of the spectrum without 
approaching too much its edges ($|V_g|\leq 2t$), $T_s$ is found to decrease as $V_g$ is increased. More precisely, we find in the large $N$ limit 
($N\gtrsim 10\xi$) that $Nk_BT_s\propto \nu_F^{-1}$ with a proportionality factor depending on $\epsilon$ (solid line in Fig.~\ref{fig_Tc}(b)). 
The temperature $T_s$ is hence given by (a fraction) of the mean level spacing at $E_F$ in this region of the spectrum ($k_BT_s\propto\Delta_F$). 
When $V_g$ is increased further, $T_s$ reaches a minimum around $|V_g|\approx 2.1t$ and then increases sharply. Outside the spectrum, this increase 
of $T_s$ with $V_g$ is well understood as follows: Since in the tunnel barrier regime, the transmission behaves (upon neglecting the disorder effect) 
as $\mathcal{T}\propto\exp(-N\zeta)$, with $\zeta=\cosh^{-1}[|E-V_g|/(2t)]$, the temperature scale below which the Sommerfeld expansion of 
integrals~\eqref{eq_coeffLi} holds is given by $k_BT_s\propto[N\left.\frac{\mathrm{d}\zeta}{\mathrm{d}E}\right|_{E_F}]^{-1}$, which yields 
$Nk_BT_s\propto t\sqrt{[(E_F-V_g)/(2t)]^2-1}$. Our numerical results are in perfect agreement with this prediction (dashed line in 
Fig.~\ref{fig_Tc}(b)).\\
\indent In Fig.~\ref{fig_Tc}(c), we investigate the behavior of $T_s$ when the spectrum of the nanowire is either scanned by varying $V_g$ at 
$E_F=0$ or by varying $E_F$ at $V_g=0$. We find that $T_s$ only depends on the part of the impurity band which is probed at $E_F$ (\textit{i.e.} 
the curves $T_s(V_g)$ and $T_s(E_F)$ are superimposed), except when $E_F$ approaches closely one edge of the conduction band of the leads. In 
that case, $T_s$ turns out to drop fast to zero as it can be seen in Fig.~\ref{fig_Tc}(c) for the case of 1D leads ($T_s\to 0$ when $E_F\to 2t$). 
This means that the divergence of the \textit{dimensionless} thermopower $S$ observed in Fig.~\ref{fig_Styp}(a) is only valid in an infinitely 
small range of temperature above $0\,\mathrm{K}$. It would be worth figuring out wether or not a singular behavior of the thermopower at the band 
edges of the conduction band persists at larger temperature.\\
\indent Let us give finally an order of magnitude in Kelvin of the temperature scale $T_s$. In Fig.~\ref{fig_Tc}(b), the lowest $T_s$ reached 
around $V_g\approx 2.1t$ is about $Nk_BT_s^{min}/t\sim 0.001$ for $\epsilon=0.004\%$. Asking for a precision of $\epsilon=1\%$ on $\mathcal{L}$, 
we get $Nk_BT_s^{min}/t\sim 0.016$. For a bismuth nanowire of length $1\,\mu\mathrm{m}$ with effective mass $m^*=0.2m_e$ ($m_e$ electron mass) 
and lattice constant $a=4.7$\,\AA, the hopping term evaluates at $t=\hbar^2/(2m^*a^2)\sim 0.84\,\mathrm{eV}$ and hence,
$T_s^{min}\sim 72\,\mathrm{mK}$. The same calculation for a silicon nanowire of length $1\,\mu\mathrm{m}$ with $m^*=0.2m_e$ and $a=$5.4\,
\AA~yields $T_s^{min}\sim 64\,\mathrm{mK}$. Those temperatures being commonly accessible in the laboratories, the results discussed in this paper 
should be amenable to experimental checks. 

%We note incidentally that our findings look in qualitative agreement with the recent experimental observation reported in Ref~\cite{Brovman2013}. 
%We stress out however that those measurements were carried out outside the low temperature coherent regime which consider, at room temperatures. 
%To describe them, inelastic effects must be included. It will be the purpose of our next paper~\cite{Bosisio2013}. 

% ******************************* CONCLUSION *********************************

\section{Conclusion}
\label{section_ccl}
We have systematically investigated the low-temperature behavior of the thermopower of a single nanowire, gradually depleted with a gate voltage 
in the field effect transistor device configuration. Disorder-induced quantum effects, unavoidable in the low-temperature coherent regime, were 
properly taken into account. We have provided a full analytical description of the behavior of the typical thermopower as a function of the gate 
voltage and have confirmed our predictions by numerical simulations. Our results show that the typical thermopower is maximized when the Fermi 
energy lies in a small region inside the impurity band of the nanowire, close to its edges. Moreover, since thermoelectric conversion strongly 
varies from one sample to another in the coherent regime, we have carefully investigated the mesoscopic fluctuations of the thermopower around 
its typical value. We have shown that the thermopower is Lorentzian-distributed inside the impurity band of the nanowire and that its fluctuations 
follow the behavior of the density of states at the Fermi energy when the gate voltage is varied. In the vicinity of the edges of the impurity band 
and outside the band, the thermopower was found Gaussian-distributed with tiny fluctuations.\\The thermopower enhancement which we predict around 
the edges looks in qualitative agreement with the recent experimental observation reported in Ref~\cite{Brovman2013}, using silicon and 
germanium/silicon nanowires in the field effect transistor device configuration. We stress out however that those measurements were carried out 
at room temperatures, and not in the low temperature coherent regime which we consider. To describe them, inelastic effects must be included. 
It will be the purpose of our next paper~\cite{Bosisio2013}. The low temperature coherent regime considered in this paper has been studied  
in Ref.~\cite{Poirier1999}, where the conductances $G$ of half a micron long Si-doped GaAs nanowires have been measured at $T=100 mK$ in the field 
effect transistor device configuration. Assuming Eq.~\eqref{eq_SeebeckMott} for evaluating the thermopower $\mathcal{S}$ from $\ln G(V_g)$, the typical 
behavior and the fluctuations of $\ln G(V_g)$ given in Ref.~\cite{Poirier1999} are consistent with the large enhancement of $\mathcal{S}$ near the band 
edges which we predict.\\
\indent Electron-electron interactions were not included in our study. A comprehensive description of the thermopower of a 1D disordered 
nanowire should definitely consider them. Nevertheless, we expect that the drastic effects of electronic correlations in 1D leading to the formation 
of a Luttinger liquid are somehow lightened by the presence of disorder. Second, the gate modulation of the thermopower we predict here is mainly 
due to a peculiar behavior of the localization length close to the edges of the impurity band. And experimentally, coherent electronic transport 
in gated quasi-1D nanowires turned out to be well captured with one-electron interference models~\cite{Poirier1999}. Of course, one could think of 
including electronic interactions numerically with appropriate numerical 1D methods but regarding the issue of thermoelectric conversion in nanowires, 
we believe the priority rather lies in a proper treatment of the phonon activated inelastic regime.\\
\indent Finally, let us discuss the potential of our results for future nanowire-based thermoelectric applications. To evaluate the efficiency of 
the thermoelectric conversion~\cite{Callen} in a nanowire, one needs to know also its thermal conductance $K$. Below the temperature $T_s$, the electron 
contribution $K_e$ to $K$ is related to the electrical conductance $G$ by the WF law. This gives $(\pi^2k_B^2 T)/(3h) [2 \exp\{-2N/\xi\}]$ 
for the typical value of $K_e$. The evaluation of the phonon contribution $K_{ph}$ to the thermal conductance of a nanowire is beyond the scope 
of the used Anderson model, since {\it static} random site potentials are assumed. In one dimension, one can expect that $K_{ph}$ 
should be also much smaller than the thermal conductance quantum $(\pi^2k_B^2 T)/(3h)$ which characterizes the ballistic phonon regime~\cite{Pendry1983,Kirczenow1998}. 
However, it remains unlikely that $K$ could be as small as $G$ for giving a large figure of merit $ZT$ in a single insulating nanowire 
at low temperature.\\
\indent Similarly, if we were to look to the delivered (electric) output power, we would find that a large length $N$ would make it 
vanish, as the electrical conductance in this regime would be exponentially small. Indeed, looking at the power factor $\mathcal{Q}=\mathcal{S}^2G$, 
which is a measure of the maximum output power~\cite{VandenBroeck2005}, we realize that the enhancement of $\mathcal{S}$ at the edge of the impurity 
band would not be enough to face the exponentially small values of $G$. Obviously, the optimization of the power factor $\mathcal{Q}$ for a single 
nanowire requires to take shorter lengths ($N\approx\xi$), while the optimization of the thermopower $\mathcal{S}$ requires to take long sizes 
($N\gg\xi$). Moreover, because of the strong variation of the localization length as the energy varies inside the impurity band, the optimization 
of the power factor for a given size $N$ requires also to not be too close from the edges of the impurity band. This illustrates the fact that a 
compromise has always to be found when thinking of practical thermoelectric applications. A way to optimize the efficiency and the output power could 
consist in taking a large array of nanowires in parallel instead of a single one. Since the conductances $G$ in parallel add while the thermopower 
$\mathcal{S}$ does not scale with the number of wires (at least if we take for $\mathcal{S}$ its typical value, neglecting the sample to sample 
fluctuations), the compromise could favor the limit of long nanowires with applied gate voltages such that electron transport occurs near the edges 
of impurity bands. Nowadays, it is possible to grow more than $10^8$ InAs nanowires~\cite{Persson2009} per $cm^2$, a large number which could balance 
the smallness of the conductance of an insulating nanowire.\\
\indent Actually, when thinking of practical applications, the results of the present paper are 
rather promising regarding Peltier refrigeration. Indeed, our conclusions drawn here for the thermopower at low temperature also hold for the Peltier 
coefficient, the two being related by the Kelvin-Onsager relation $\Pi=\mathcal{S}T$. One could imagine to build up Peltier modules with doped nanowires 
for cooling down a device at sub-Kelvin temperature in a coherent way. Besides, whether it be for energy harvesting or Peltier cooling, it would be 
worth considering more complicated setups using the nanowire as a building block (e.g. arrays of parallel nanowires in the field effect transistor 
device configuration) in order to reach larger values of output electric/cooling power. 

\acknowledgments{Useful discussions with G. Benenti, O. Bourgeois, C. Gorini, Y. Imry, K. Muttalib and H. Shtrikman are gratefully acknowledged. This work has been supported by CEA through the DSM-Energy Program (project E112-7-Meso-Therm-DSM).}

% ******************************* APPENDIX *********************************

\appendix
\section{Self-energy of the 2D leads} 
\label{app_SelfNRJ}
We give here the expression of the retarded self-energy of a 2D lead (made of a semi-infinite square lattice with hopping term $t$) connected 
at one site (with coupling $t$) to a nanowire of $N$ sites length. It is a $N\times N$ matrix $\Sigma$ with only one non-zero component denoted 
$\sigma$. To calculate $\sigma$, we calculate first the retarded Green's function of an infinite square lattice~\cite{Economou2006} and then 
deduce with the method of mirror images the retarded Green's function of the semi-infinite 2D lead~\cite{Molina2006}, that we evaluate at the site 
in the lead coupled to the nanowire to get $\sigma$. Analytic continuations of special functions are also required, they can be found for example 
in Ref.~\cite{Morita1971}. Introducing the notation $z=E/(4t)$, we find for $\sigma=\mathrm{Re}(\sigma)+i\,\mathrm{Im}(\sigma)$ 
\begin{align}
\mathrm{Re}(\sigma) &= tz \pm \frac{2t}{\pi}\left[ \mathcal{E}(z^2)-(1-z^2)\mathcal{K}(z^2)\right] \label{eq_app_Sigma2Dlead1}\\
\mathrm{Im}(\sigma) &= \frac{2t}{\pi}\left[ -\mathcal{E}(1-z^2)+z^2\mathcal{K}(1-z^2)\right]
\end{align}
with a $+$ sign in Eq.~\eqref{eq_app_Sigma2Dlead1} when $-4t\leq E\leq 0$ and a $-$ sign when $0\leq E\leq 4t$. If the energy $E$ is outside 
the conduction band of the lead ($|E|>4t$), we get
\be
\sigma=tz\left[1-\frac{2}{\pi}\mathcal{E}\left(\frac{1}{z^2}\right)\right]\,.
\ee
In the three above equations, $\mathcal{K}$ and $\mathcal{E}$ stand for the complete elliptic integrals of the first and second kind respectively. 
They are defined as
\begin{align}
\mathcal{K}(z) &= \int_0^{\pi/2}\!d\phi\,[1-z\sin^2\phi]^{-1/2} \\
\mathcal{E}(z) &= \int_0^{\pi/2}\!d\phi\,[1-z\sin^2\phi]^{1/2}\,.
\end{align}

\section{Thermopower of a clean tunnel barrier} 
\label{appThermopowerCTB}
In this appendix, we derive Eq.~\eqref{eq_S0TB}. We consider a clean nanowire with on-site potentials $V_g$, connected via its extreme sites 
$1$ and $N$ to two identical semi-infinite leads. In order to investigate the tunnel barrier regime, we assume that the energy $E$ of the 
incoming electrons lies outside the spectrum $[-2t,2t]$ of the nanowire. Let us say that $E\geq V_g+2t$ to fix the ideas. In the basis 
$\{1,N,2,...,N-2\}$, the retarded Green's function $G=[E-\mathcal{H}-\Sigma_L-\Sigma_R]^{-1}$ of the system reads
\be
G=
\begin{pmatrix}
A & B \\
\tilde{B} & C
\end{pmatrix}^{-1}
\ee
where \textit{(i)} $A=(E-V_g-\sigma)\bf{1}_2$ ($\bf{1}_2$ being the $2\times 2$ identity matrix and $\sigma$ the non-vanishing element of 
$\Sigma_L$ and $\Sigma_R$), \textit{(ii)} $B$ [$\tilde{B}$] is a $2\times (N-2)$ [$(N-2)\times 2$] matrix with all zero components except 
two equal to $t$ coupling the sites $1$ and $N$ to their neighbors $2$ and $N-1$, and \textit{(iii)} $C$ is a $(N-2)\times(N-2)$ symmetric 
tridiagonal matrix with all diagonal elements equal to $E-V_g$ and all elements on the first diagonals below and above the main one equal 
to $t$. Using the Fisher-Lee formula~\eqref{eq_FisherLee}, we write the transmission function $\mathcal{T}(E)$ as 
\begin{align}
\mathcal{T}(E)&=\mathrm{Tr}\left[
\begin{pmatrix}
\gamma & 0 \\
0 & 0
\end{pmatrix}
G_A
\begin{pmatrix}
0 & 0 \\
0 & \gamma
\end{pmatrix}
G_A^\dagger
\right] \\
     & =\gamma^2|G_A^{(1N)}|^2   \label{eq_app_Transmission}
\end{align}
where $G_A$ is the $2\times 2$ submatrix in the top left-hand corner of $G$, $G_A^{(1N)}$ its top right element and $\gamma=-2\mathrm{Im}(\sigma)$. 
To calculate $G_A$, we first notice that
\be
\label{eq_app_GA}
G_A=(A-BC^{-1}\tilde{B})^{-1}=(A-t^2C_{\square}^{-1})^{-1}
\ee
where $C_{\square}^{-1}$ is a $2\times 2$ submatrix of $C^{-1}$ made up of the four elements located at its four corners. Second, we make use 
of Ref.~\cite{Hu1996} for computing the inverse of the symmetric tridiagonal matrix $C$. We get
\be
C_{\square}^{-1}=
\begin{pmatrix}
\alpha & \beta \\
\beta & \alpha
\end{pmatrix}
\ee
with
\begin{align}
\alpha &=-\frac{\cosh(\zeta)\cosh((N-2)\zeta)}{t\sinh(\zeta)\sinh((N-1)\zeta)} \\
\beta &=-\frac{\cosh(2\zeta)+(-1)^{N-1}}{2t\sinh(\zeta)\sinh((N-1)\zeta)}   \label{eq_app_beta}
\end{align}
and $\zeta=\cosh^{-1}[(E-V_g)/(2t)]$. Plugging Eqs.~(\ref{eq_app_GA}-\ref{eq_app_beta}) into Eq.~\eqref{eq_app_Transmission}, we deduce the 
exact transmission function $\mathcal{T}(E)$, and hence the thermopower $S$ defined by Eq.~\eqref{eq_df_S}. An expansion at large $N$ yields 
$\mathcal{T}\propto\exp(-2N\zeta)$ (as expected for a tunnel barrier) and the expression~\eqref{eq_S0TB} for the thermopower. The same 
demonstration can be made for the energy range $E\leq V_g-2t$.

\bibliography{Thermo1dchain}

\end{document}